\documentclass[a4paper,11pt]{article}
\pdfoutput=1 

\usepackage{jinstpub} 

\usepackage{titlesec}

\title{A flexible and low-cost open-source IPMC mezzanine for ATCA boards based on OpenIPMC}

\author[a,1]{L. Calligaris,\note{Corresponding author.}}
\author[a]{A. Cascadan}
\author[b]{L. E. Ardila-Perez,}
\author[c]{M. Pesaresi}
\author[c]{G. Fedi}
\author[d]{A. Peck}
\author[d]{D. Gastler}

\affiliation[a]{Scientific Computing Center (NCC) of São Paulo State University (UNESP),\\, Rua Dr. Bento Teobaldo Ferraz, 271, São Paulo - SP, 01140-070, Brazil}
\affiliation[b]{Institute for Data Processing and Electronics (IPE) of Karlsruhe Institute of Technology,\\Hermann-von-Helmholtz-Platz 1, D-76344 Eggenstein-Leopoldshafen, Germany}
\affiliation[c]{Imperial College London, Physics Department,\\Blackett Laboratory, Prince Consort Rd, London, SW7 2BW, UK}
\affiliation[d]{Department of Physics at Boston University,\\590 Commonwealth Avenue, Boston, MA 02215, US}

\emailAdd{luigi.calligaris@cern.ch}

\abstract{
This work presents the development of an Intelligent Platform Management Controller mezzanine in a Mini DIMM form factor for use in electronic boards compliant to the PICMG Advanced Telecommunication Computing Architecture (ATCA) standard. The module is based on an STMicroelectronics STM32H745 microcontroller running the OpenIPMC open-source software. The mezzanine has been successfully tested on a variety of ATCA boards being proposed for the upgrade of the experiments at the HL-LHC, with its design and firmware being distributed under open-source hardware license.
}


\proceeding{Topical Workshop on Electronics for Particle Physics - TWEPP2021\\
  September 20-24, 2021}

\begin{document}
\maketitle
\flushbottom

\section{Introduction}
\vspace{-3mm}
The use of electronic boards compliant to the PICMG Advanced Telecommunication Computing Architecture (ATCA) standard \cite{picmg_3_0} is being pursued in a number of development projects targeting the upgrades of the Large Hadron Collider (LHC) experiments. Examples of ATCA boards designed for use in future upgrades of LHC experiments are the Apollo \cite{Albert:2019kvt}, Serenity \cite{Rose:2019oiy}, Pulsar-IIb \cite{Ajuha:2017frj,Okumura:2013ych} and CMS DAQ and Timing Hub \cite{Hegeman:2725211} boards. In the context of the ATCA standard Hardware Platform Management (HPM) infrastructure, the management of each Field Replaceable Unit (FRU), such as an electronic board, is delegated to an on-board Intelligent Platform Management Controller (IPMC), which can be installed as a permanent part of the board or as an add-on mezzanine to coordinate the FRU operation with the Shelf Manager Controller (ShMC). Various IPMC mezzanines have been proposed for use in the upgrades of LHC experiments, beginning with the IPMC mezzanine designed by the \emph{Laboratoire d'Annecy de Physique des Particules} (LAPP) \cite{lapp-ipmc}, which established a JEDEC MO-244 MiniDIMM form factor as a common standard for these mezzanines, followed by the Pulsar-IIb IPMC \cite{ipmc-pulsar2b_ramalho_ieee_talk,ipmc-pulsar2b_paiva_thesis}, the CERN IPMC \cite{cern-ipmc} and the ZYNQ-IPMC \cite{zynq-ipmc}. None of the aforementioned IPMC mezzanines offers at the same time a solution which is hardware-unspecific and customizeable for any ATCA board, free and open-source, compliant with the LAPP IPMC pinning and fitting in a vertical DIMM socket - as used in the ATCA boards of interest - without invading the volume reserved to a nearby ATCA slot. A previous publication presented OpenIPMC \cite{openipmc-sw, openipmc_repo}, an architecture-independent, free and open-source software written in embedded C language and based on the free and open-source real-time operating system FreeRTOS\cite{freertos}, implementing the functionality required for an IPMC. Following the initial development targeting Zynq Ultrascale+ devices \cite{atca-zynqmp-ipmc}, OpenIPMC has been ported to STM32 devices. This publication describes the development of an open-source, low-cost MiniDIMM IPMC mezzanine running OpenIPMC, designed to act as a drop-in replacement for existing IPMC mezzanines. The open-source nature of the board allows for modifications, customzations and fixes in its firmware and hardware by third parties independent from the original developers, which is a strong point in favour of long-term serviceability, and relieves the users from the need to sign non-disclosure agreements, as in the case of the commercial, closed-source IPMC solutions.

\section{Mezzanine hardware: OpenIPMC-HW}
\begin{figure}[htbp]
    \centering
    \includegraphics[width=0.49\linewidth]{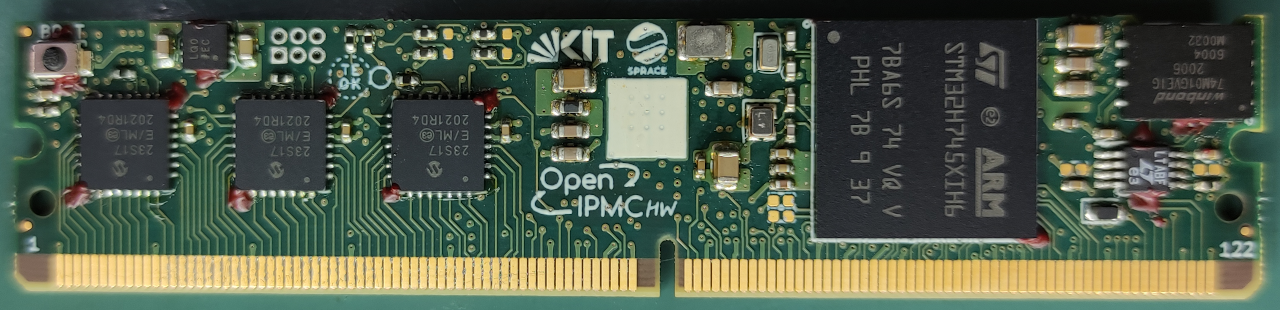}
    \includegraphics[width=0.49\linewidth]{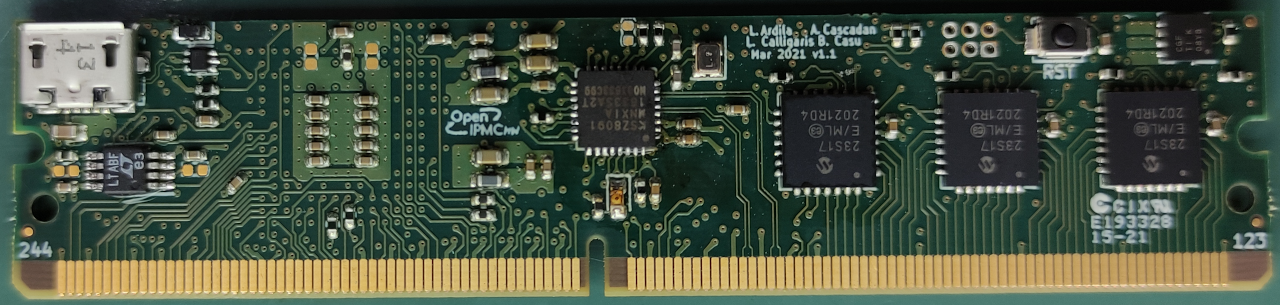}
    \caption{\label{fig:openipmc-hw_v1.1_pcb_front_back} Front and back view of the OpenIPMC-HW mezzanine, hardware version 1.1.}
\end{figure}

The mezzanine presented in this work has been named OpenIPMC-Hardware (OpenIPMC-HW) \cite{openipmc_repo}, to reflect its origins rooted in the design of the OpenIPMC software, and it has been designed using KiCad, a free and open source Electronic Design Automation (EDA) software \cite{kicad}. The heart of the mezzanine is a powerful STM32H745XIH6 \cite{stm32h745xig} microcontroller unit (MCU) by STMicroelectronics in a TFBGA240 package, which features a main ARM Cortex-M7 core capable of running at 480 MHz, an auxiliary ARM Cortex-M4 core capable of running at 240 MHz, 2 MiB of Flash memory and 1 MiB of SRAM, 4 I$^{2}$C, 4 USART, 5 UART, 6 SPI and Ethernet MAC hard peripherals, among others. As required by the PICMG standard, the I2C buses dedicated to IPMB communication are protected by suitable fail-safe Linear Technology LTC4300-1 I2C buffers. Front and back views of the assembled mezzanine are shown in figure \ref{fig:openipmc-hw_v1.1_pcb_front_back}.

The board is manufactured as a 82.00 mm x 18.30 mm, 1.0 mm-thick, 8-layer, ENIG-finished PCB in traditional FR-4 material with through-hole vias, and hard gold plated contacts with 45° chamfer on the card edge connector. The small dimensions of the board and the relative low speed of the signals involved in the operation of the mezzanine (the Ethernet connection having the fastest signals) do not require the adoption of special impedance control measures in board manufacturing. The 240-ball MCU package provides plenty of GPIO capability, allowing to service most of the card edge connector pins using the microcontroller pins and dispensing from the need of an on-board FPGA. A subset of 90 card-edge connector pins, dedicated to the implementation of slow control signals for the operation of add-on PICMG Advanced Mezzanine Cards (AMC) \cite{picmg_amc.0} is indirectly operated by the MCU through a set of six Microchip MCP23S17-ML SPI-controlled GPIO expanders. The mezzanine supports the Ethernet (ETH) capability of the MCU through the use of a Micrel/Microchip KSZ8091MNX Ethernet PHY, compatible with the 10BASE-T/100BASE-TX Ethernet standard, connected to the MCU via a Media-Independent Interface (MII) bus. A WSON-8 landing pattern on the PCB allows the mounting of a QSPI Flash memory, such as the Winbond W25Q01JVZEIM and W74M01GVZEIG, which is used in device configuration and in the firmware update mechanism. A Micro-USB B socket allows to power the board from a traditional USB cable connected to a PC, offering access to the USB PHY hardware of the microcontroller to update the firmware and to get access to a shell terminal, thanks to compatibility of the MCU with the USB Device Firmware Upgrade (DFU) and Communication Device (CDC) classes, dispensing from the need of using a hardware-specific debug adapter for basic firmware maintenance and configuration.

\begin{figure}[htbp]
    \centering
    \includegraphics[width=0.50\linewidth]{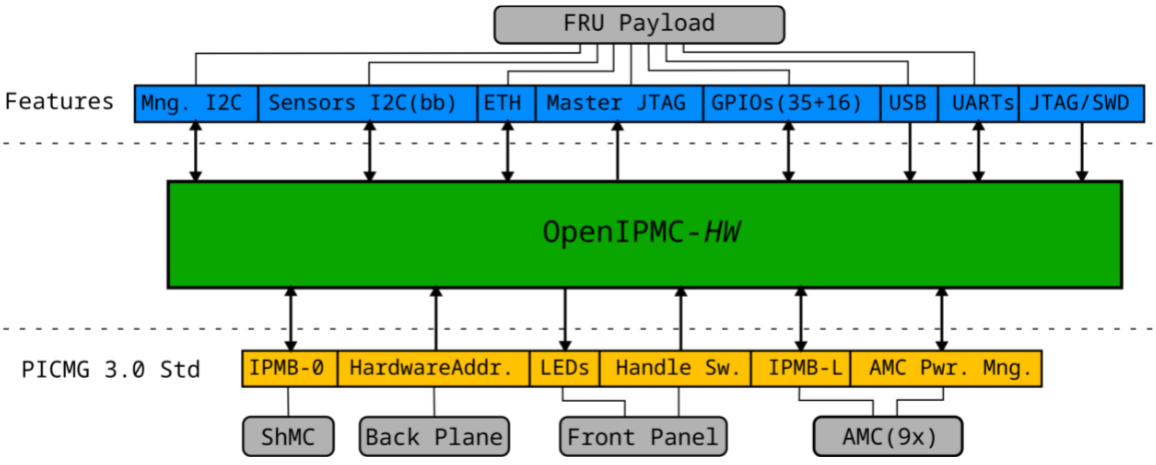}
    \caption{\label{fig:openipmc-hw_carrier_dimm_interface.png} Schematic of the card edge interface between OpenIPMC-HW and the ATCA carrier board. The bottom signals are mandated by the PICMG standard\cite{picmg_3_0}, the top ones are optional features provided by OpenIPMC-HW.}
\end{figure}

The card edge connector pin assignment of the mezzanine is designed to be pin-compatible with the currently agreed pin standard for MiniDIMM IPMC mezzanines, as agreed in the xTCA Interest Group and SoC Interest Group meetings at CERN, including the debugging UART channel between the IPMC and the ATCA carrier board front panel, the one between the IPMC and the management SoC for remote access to the SoC Linux terminal, and the one between the SoC and the IPMC used for boot-up configuration of the former. Figure \ref{fig:openipmc-hw_carrier_dimm_interface.png} summarizes the various signals carried across the card edge connector. Overall the active board components are inexpensive, costing a few dollars each and with the most expensive one being the MCU, priced around USD 17 at the time of prototype manufacturing at the end of 2020.

\section{Mezzanine firmware: OpenIPMC-FW}
\begin{figure}[htbp]
    \centering
    \includegraphics[height=30mm]{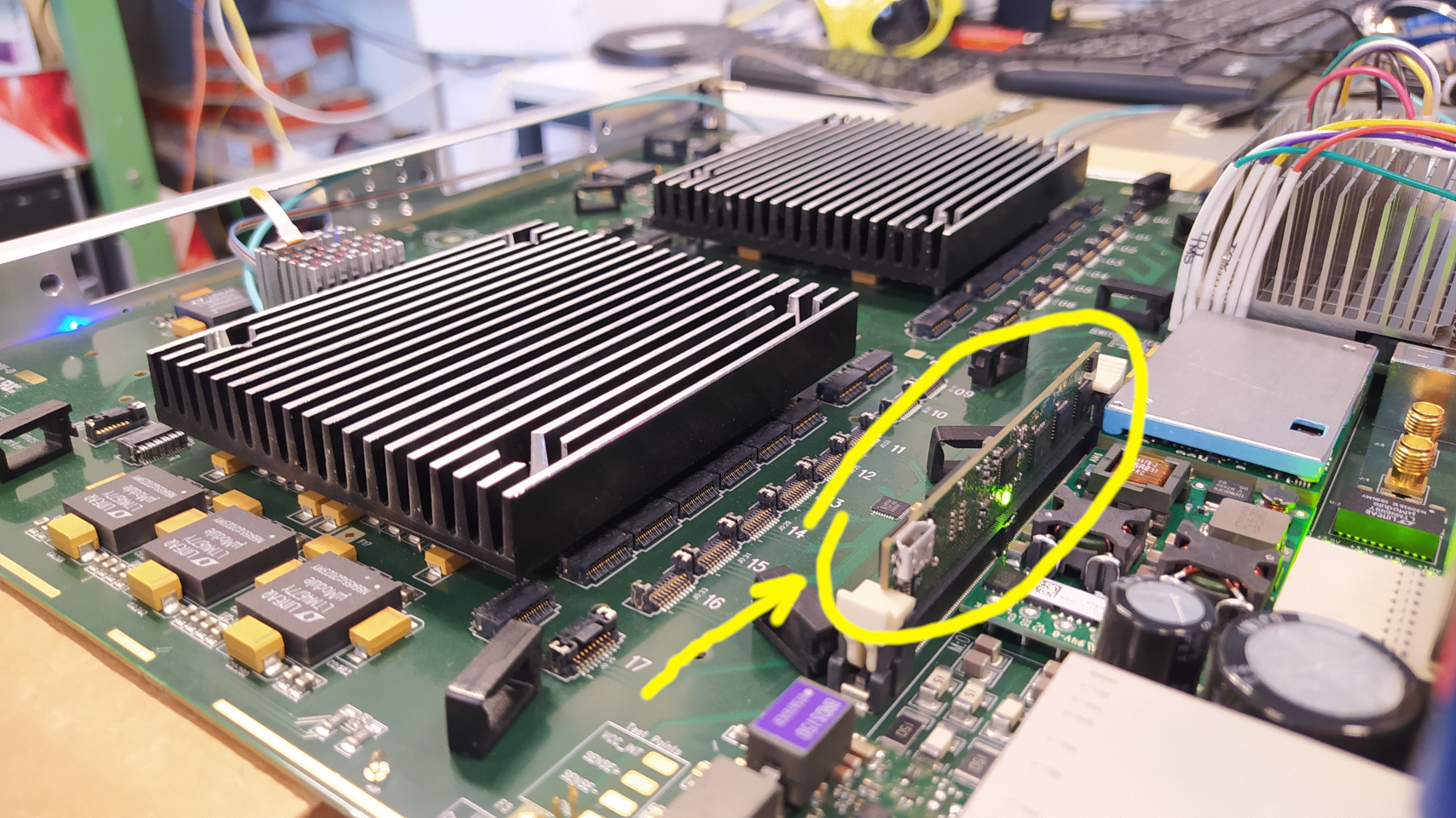} \hspace{15mm}
    \includegraphics[height=30mm]{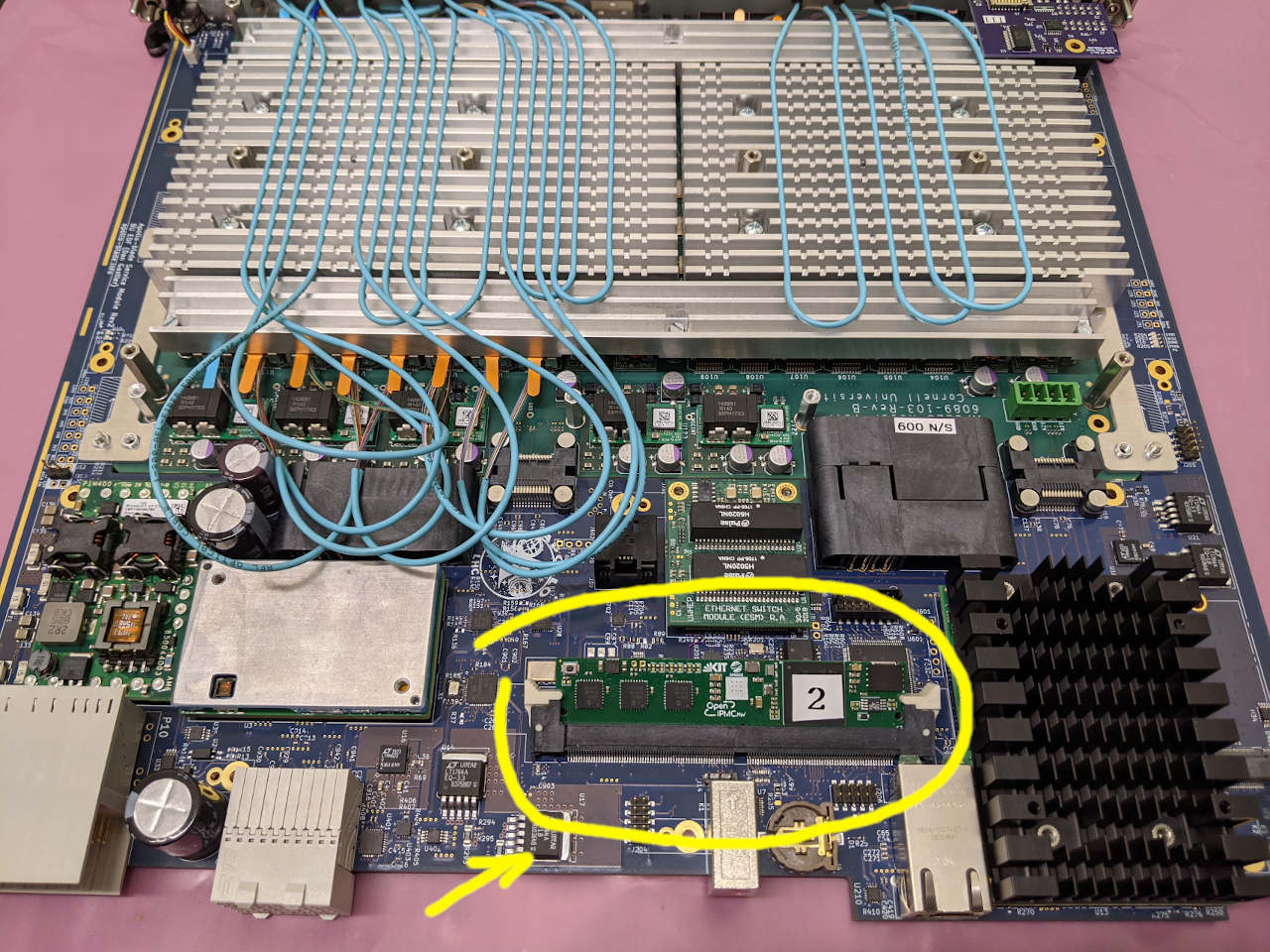}
    \caption{\label{fig:openipmc-hw_in_serenity_apollo_scale_arrow} OpenIPMC-HW running in the Serenity (left) and Apollo (right) boards during development.}
\end{figure}

The mezzanine firmware is based on the STMicroelectronics-provided customization of FreeRTOS for STM32H7 microcontrollers, in which OpenIPMC runs alongside other services like the lwIP network stack \cite{lwip}, a command line interface server which can be accessed via USB-CDC, Telnet and UART. So far a PICMG HPM.1-compliant firmware update process has been integrated into the firmware and a Xilinx Virtual Cable server is planned to be integrated soon, to support the programming and debug of FPGAs and Systems-on-a-Chip installed on the carrier board. Following the design workflow expected for OpenIPMC board implementations, two customizations of the firmware have been in development so far - one targeting the Serenity board and one targeting the Apollo board, shown in figure \ref{fig:openipmc-hw_in_serenity_apollo_scale_arrow} - such that the OpenIPMC board-specific controls for each of the boards can be tailored to its specific needs (e.g. to match the correct sequence of signals exchanged with the carrier management SoC during board boot-up).

In the case of the Serenity board, this firmware customization work involved the developers of OpenIPMC, while in the case of the Apollo the OpenIPMC developers have been largely excluded from this process on purpose, to test the development experience from the point of view of an OpenIPMC user. In both cases the development has progressed smoothly, and the feedback received from the work on Apollo has been very valuable in improving the OpenIPMC API.

\section{Testing}
\begin{figure}[htbp]
  \centering
  \includegraphics[height=30mm]{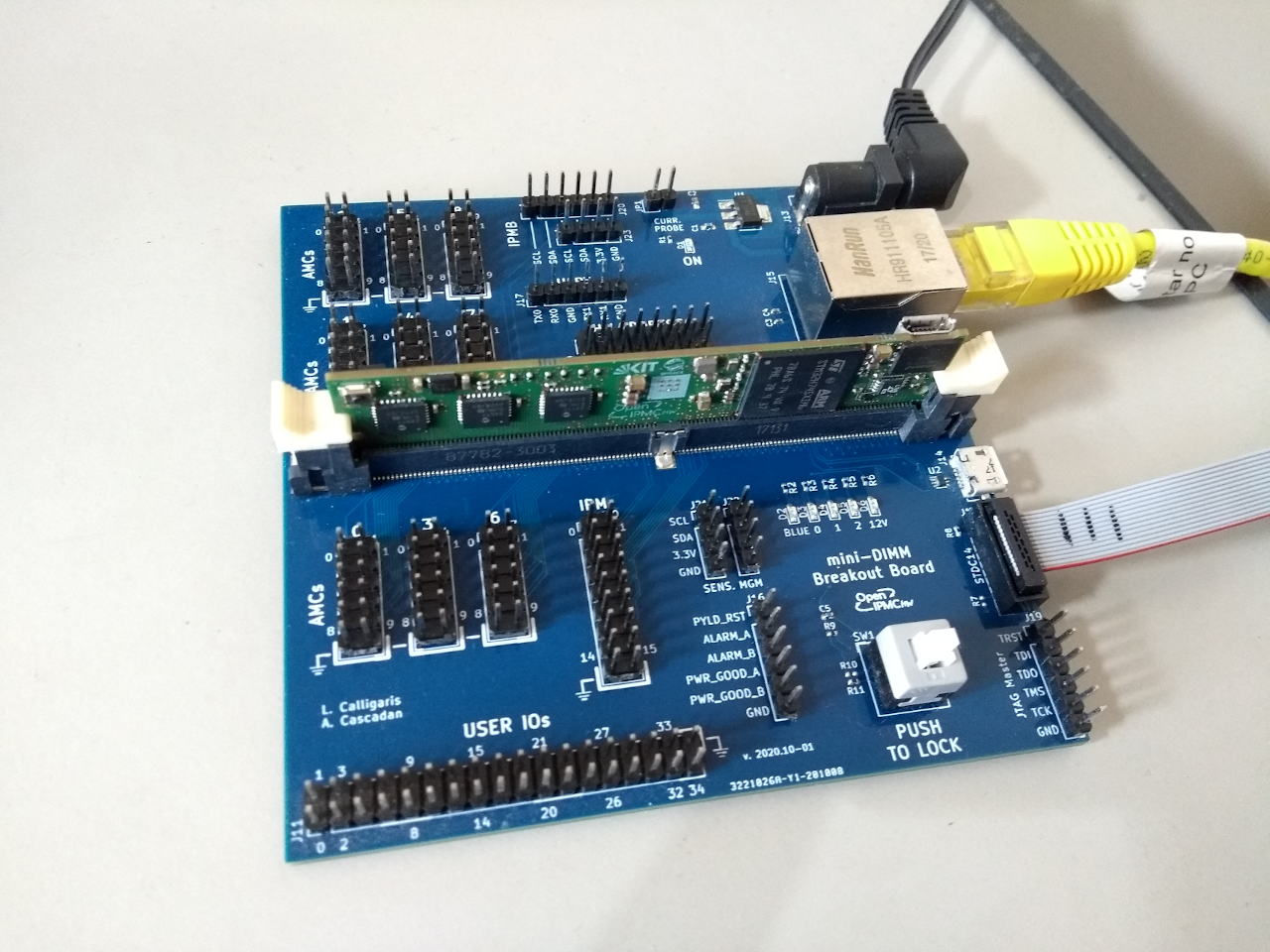}
  \hspace{15mm}
  \includegraphics[height=30mm]{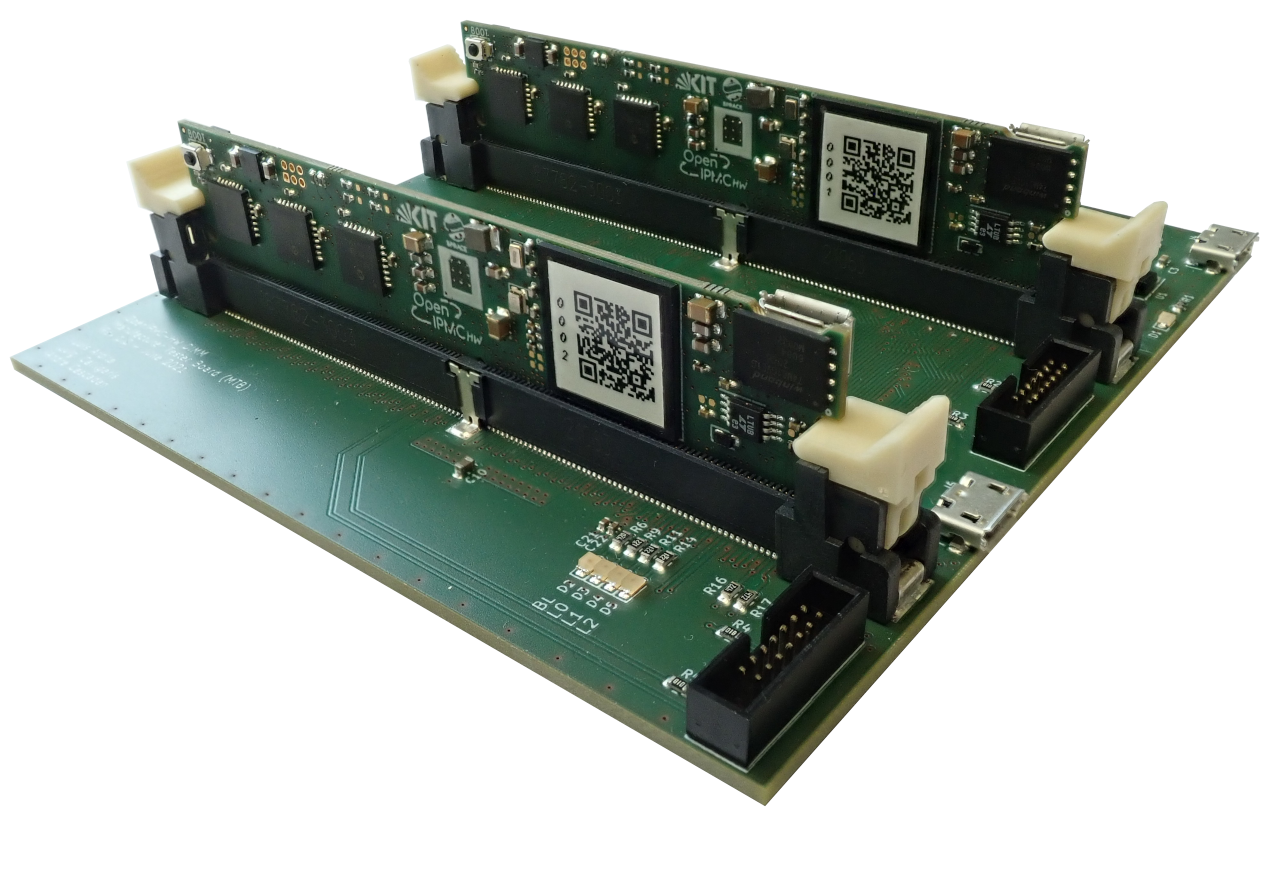}\\
  \caption{\label{fig:Breakout-MTB} Breakout Board hosting an OpenIPMC-HW mezzanine (left) and  Manufacturing Test Board hosting two mezzanines, one of them acting as test master, during post-manufacturing testing (right).}
\end{figure}

Initial development of the firmware and testing of OpenIPMC-HW were carried out on a custom-designed breakout board, shown on the left in figure \ref{fig:Breakout-MTB}. This board provides stabilized power to the mezzanine, exposes the mezzanine Ethernet signals to an 8P8C connector, the USB bus to a Micro-USB connector, the slave JTAG/SWD signals to an STDC14 connector for connection of the MCU with a STLinkV3 debug adapter and the remainder of the card edge connector signals to standard 100-mil headers, which can be connected to external devices using jump wires. Alongside the breakout board, a Manufacturer Tester Board (MTB) has been designed to provide a simple mechanism to test new boards following manufacturing. This board wires the pins of the mezzanine being tested against the ones of a mezzanine programmed with a dedicated test firmware, which can be controlled by a PC through a USB-CDC serial terminal to drive the tests.

\begin{figure}[htbp]
  \centering
  \includegraphics[height=35mm]{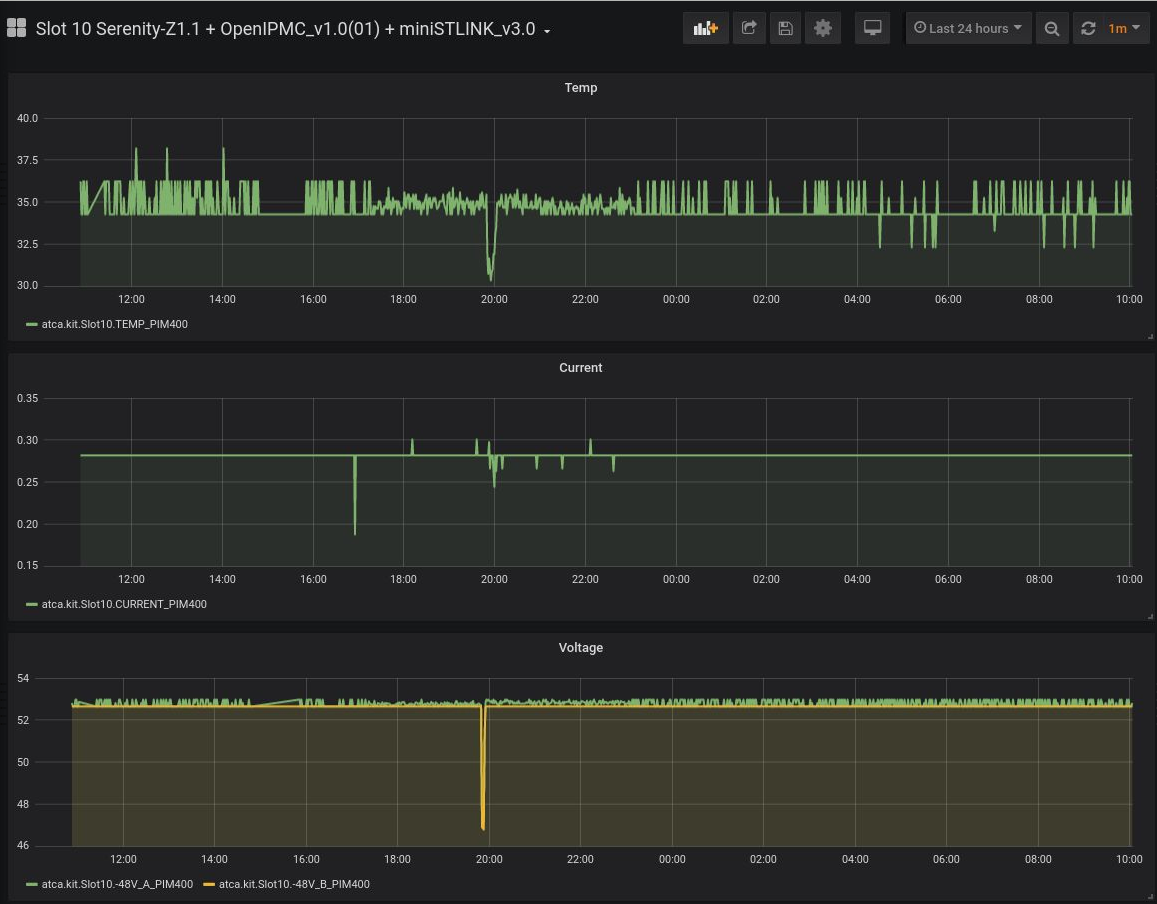}
  \hspace{15mm}
  \includegraphics[height=35mm]{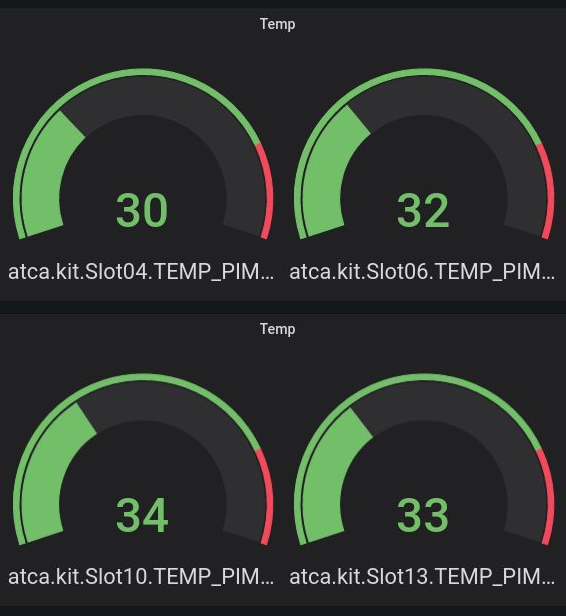}\\
  \caption{\label{fig:CarbonGrafana} The Carbon/Grafana server showing IPMI data reported by OpenIPMC-HW. (right) temperature of the Power Input Module (PIM400) for four Serenity boards (left: top, middle, bottom) time series plot of temperature, current and voltage for a specific board.}
\end{figure}

\vspace{-1.5mm}
The firmware was initially adapted trivially for basic operation in the Pulsar-IIb board for testing on the setup in São Paulo, followed by tests and development on Serenity boards on the test setups at KIT and CERN, and on Apollo boards at Boston University. The compliance of the OpenIPMC-HW/Serenity setup with the PICMG standards was so far tested three times at CERN using a commercial automated testing software developed by Polaris Networks, which proved valuable in directing the firmware development efforts. In those automated tests, performed the last time in May 2021, the IPMC showed a coverage of 55\% against the ATCA standard. Since then, changes to the firmware were implemented to improve standards coverage and a new series of compliance tests is expected for late 2021. An online sensor monitoring system, based on a Carbon/Grafana server pair reading out the values of the sensors published by OpenIPMC-FW on the ShMC, has been set up at CERN and KIT to analyze the long-term behavior of the IPMC mezzanines under test with plots and dashboards, such as those shown in figure \ref{fig:CarbonGrafana}. With this setup, we ran several OpenIPMC-HWs for extended amounts of time (order of months) at the three test sites, helping in the improvement of minor issues and at the same time showing a very good overall reliability.

\section{Summary and outlook}
This work presented OpenIPMC-HW, an IPMC mezzanine in a Mini-DIMM form factor running OpenIPMC and based on an STM32. The board is being developed targeting its use in Serenity and Apollo ATCA carrier boards, but its scope could be extended to other ATCA hardware. The board firmware is actively being developed and tested for compliance against the ATCA standard, with new features expected to be finalized in the near future, such as support for the PICMG HPM.1 firmware upgrade and the inclusion of an XVC server in the firmware.

\acknowledgments
The authors acknowledge the Fundação de Amparo à Pesquisa do Estado de São Paulo for its financial support through grants number 18/18955-0 and 18/25225-9. We thank the members of the CMS Phase-2 Tracker Upgrade Data Processing Systems group for the exchange of ideas, and in particular Gregory Iles, Eric Hazen and Peter Wittich for the help in defining the requirements for IPMCs used in the Phase-2 back-end boards. We want to thank the CERN EP-ESE group, in particular Julian Maxime Mendez, for the access to the Polaris ACTA compliance tester at CERN.

\renewcommand*{\bibfont}{\normalfont\fontsize{6}{4}\selectfont}

\end{document}